%% file: main.tex
\documentclass{article}
\pdfoutput=1
\usepackage{amsmath, bm, graphicx, mlspconf, glossaries}
\usepackage{blindtext}
\usepackage{hyperref}
\usepackage{afterpage}
\usepackage{multicol}
\usepackage{float}
\usepackage{epstopdf}
\usepackage{xcolor}
\usepackage{amsfonts}

\usepackage{subcaption} 

\makeatletter
\renewcommand{\section}{\@startsection{section}{1}{\z@}%
  {-2.5ex \@plus -0.5ex \@minus -.2ex}%
  {1.50ex \@plus.2ex \@minus-.2ex}%
  {\normalfont\large\bfseries\centering}}

\renewcommand{\subsection}{\@startsection{subsection}{2}{\z@}%
  {-1.50ex \@plus -0.5ex \@minus -.2ex}%
  {0.50ex \@plus .2ex}%
  {\normalfont\normalsize\bfseries}}

\makeatother

\toappear{2024 IEEE International Workshop on Machine Learning for Signal Processing, Sept.\ 22--25, 2024, London, UK}

\copyrightnotice{979-8-3503-2411-2/23/\$31.00 {\copyright}2023 IEEE}

\title{SPEED: Scalable Preprocessing of EEG Data for Self-Supervised Learning}

\name{%
    \fontsize{10.5pt}{11pt}\selectfont
    \begin{tabular}{c}
    Anders Gjølbye$^{\star}$ \qquad  Lina Skerath$^{\star}$\\
    William Lehn-Schiøler$^{\star}$ \qquad Nicolas Langer$^{\dagger}$ \qquad Lars Kai Hansen$^{\star}$
    \end{tabular}
    \vspace{-3pt}
    \thanks{This work is supported by The Pioneer Centre for AI, DNRF grant number P1, The Novo Nordisk Foundation grant NNF22OC0076907 "Cognitive spaces - Next generation explainability".}
}

\address{%
    \small
    \begin{tabular}{c}
    \begin{tabular}{@{}c@{}}
    $^{\star}$Technical University of Denmark \\
    Department of Applied Mathematics and Computer Science \\
    2800 Kgs. Lyngby, Denmark
    \end{tabular}
    \qquad \qquad
    \begin{tabular}{@{}c@{}}
    $^{\dagger}$University of Zürich \\
    Department of Psychology \\
    8050 Zurich, Switzerland
    \end{tabular}
    \end{tabular}
    \vspace{-11pt}
}

\begin{document}
\ninept

\maketitle

\begin{abstract}
Electroencephalography (EEG) research typically focuses on tasks with narrowly defined objectives, but recent studies are expanding into the use of unlabeled data within larger models, aiming for a broader range of applications. This addresses a critical challenge in EEG research. For example, Kostas et al. (2021) show that self-supervised learning (SSL) outperforms traditional supervised methods. Given the high noise levels in EEG data, we argue that further improvements are possible with additional preprocessing. Current preprocessing methods often fail to efficiently manage the large data volumes required for SSL, due to their lack of optimization, reliance on subjective manual corrections, and validation processes or inflexible protocols that limit SSL. We propose a Python-based EEG preprocessing pipeline optimized for self-supervised learning, designed to efficiently process large-scale data. This optimization not only stabilizes self-supervised training but also enhances performance on downstream tasks compared to training with raw data.

\end{abstract}

\begin{keywords}
Electroencephalography (EEG), Data Preprocessing, Self-Supervised Learning
\end{keywords}

\input{Sections/1_Introduction}
\input{Sections/2_Methods}

\input{Sections/3_Results}
\input{Sections/4_Conclusion}

\bibliographystyle{IEEEbib}
\bibliography{main}

\end{document}

%% file: Sections/1_Introduction.tex
\section{Introduction} \label{sec:intro}

As the processing of EEG measurements becomes more sophisticated, the range of applications expands. Much EEG-based modelling is focused on limited-scale, specific tasks, with fewer large-scale projects targeting broader applicability. For limited-scale tasks, preprocessing is often a core dimension, making modelling of small and noisy datasets possible. Such methods have not yet reached a level of performance that excels across various tasks like in the case of large language models (LLM). Nevertheless, some inspiration is still drawn from LLMs, such as the application of self-supervised learning (SSL) that makes use of unlabeled data. The most prominent results have been achieved by Kostas et al. \cite{kostas2021bendr}. Their BENDR model shows that while not effective on their own, SSL can improve downstream task performance. The authors claim that deep learning is sufficient to learn even from raw data, yet there is an important scientific question: Does preprocessing help SSL? This is the question we set out to test in the present work.

\quad Current preprocessing methods do not scale to the substantial volume of data required for SSL. While current pipelines allow for manual correction and validation of data, this makes them subjective and challenges reproducibility. Additionally, the correction process is often too task-specific for SSL, given the diverse nature of downstream tasks. Common preprocessing methods also tend to lead to significant data loss, especially considering that the largest available EEG dataset, the Temple University Hospital EEG Corpus (TUEG) \cite{obeid2016temple}, comprises EEG of highly variable signal-to-noise. Most approaches become unfeasible with terabytes of data, underlining the need for robust, optimised pipelines capable of efficiently handling such large volumes.
To address these challenges, this paper introduces SPEED: \textbf{S}calable \textbf{P}reprocessing for \textbf{EE}G \textbf{D}ata, a Python-based large-scale EEG data preprocessing pipeline tailored for self-supervised learning. Our proposed pipeline is optimized for massive data processing, efficiently leveraging hardware. We provide both pretraining and downstream procedures while avoiding removing important information from the signal. We aim to establish standardized procedures and enhance accessibility.

\quad Building upon existing frameworks, we maintain consistency while introducing additional functionalities and improvements, motivated by experiments. We demonstrate that preprocessing data for SSL models increases contrastive accuracy and yields increased performance in downstream tasks compared to training on raw data. Additionally, we address the challenges of preprocessing the massive TUEG dataset by providing 
comprehensive log files to enable reproducibility and further work with cleaned data.


\quad All code used in this research, along with references to the datasets, are made publicly accessible. \footnote{\url{https://github.com/AndersGMadsen/SPEED}}.

\section{Background} 



EEG data varies among subjects, hardware, and environmental factors and is susceptible to noise and artifacts. Therefore, a lot of research in this field focuses on signal cleaning to 
extract the genuine brain signal.
Historically, MATLAB has been the dominant platform for EEG research, leading to the development of numerous frameworks and pipelines for EEG analysis. EEGLAB \cite{delorme2004eeglab}, Brainstorm \cite{tadel2011brainstorm} and FieldTrip \cite{oostenveld2011fieldtrip} 
are some of the most widely used frameworks. 
Other projects, such as the PREP pipeline \cite{PREP}, aim to standardize larger-scale EEG preprocessing, incorporating advanced techniques for removing line noise and performing robust average referencing through iterative detection and interpolation of artifact-contaminated channels. 
The Automagic pipeline \cite{automagic} is designed to aggregate currently available preprocessing methods and provide objective standardized quality assessment techniques. 

\quad 
 Based on Python's open-source nature and widespread accessibility, MNE \cite{gramfort2013meg} offers a comprehensive suite of tools. Other libraries and tools for bad channel detection were developed or adapted from MATLAB, such as FASTER \cite{nolan2010faster}, Autoreject \cite{jas2017autoreject}, and the adapted PREP pipeline. However, available pipelines cannot handle the massive amounts needed for SSL and are not designed for SSL objectives.

\quad Kostas et al.\ \cite{kostas2021bendr} and Banville et al.\ \cite{banville2021uncovering} find self-supervision to outperform or reach competitive performance compared to supervised methods. Banville et al.\ also find interpretable latent structures in the learned embeddings, highlighting the efficacy of SSL in capturing EEG-specific properties. However, they also perform minimal processing using the same TUEG dataset. In their NEURO-GPT model, Cui et al.\ \cite{cui2023neuro} preprocess their data using the Brainstorm software performing simple bad channel detection, interpolation, and filtering. We suggest a similar but more comprehensive amount of preprocessing, showing that preprocessing biases models to even clearer structures and better generalization and downstream results.

\begin{figure*}[h!]
    \centering
    \includegraphics[width=\textwidth]{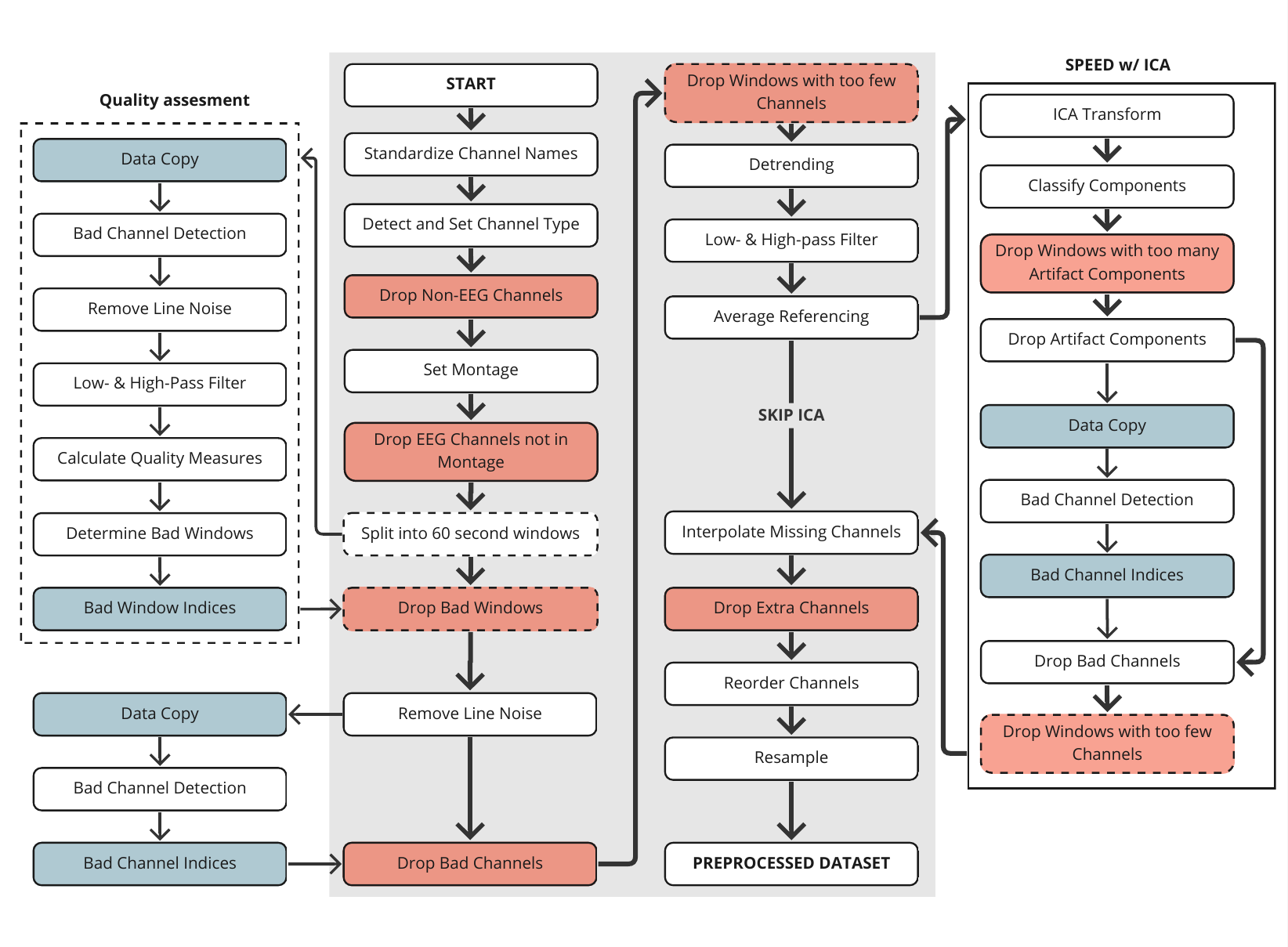}
    \vspace{-18pt}
    \caption{An overview of the \textbf{S}calable \textbf{P}reprocessing for \textbf{EE}G \textbf{D}ata (SPEED) pipeline and the SPEED w/ICA pipeline. These pipelines are crafted to manage massive amounts of EEG data and are intended for use with a self-supervised learning (SSL) model. The red colour highlights removing faulty or unnecessary data, while dotted lines signify pipeline components that are excluded when preprocessing downstream datasets.}
    \label{fig:pipeline}
\end{figure*}

%% file: Sections/2_Methods.tex
\section{Data}

\subsection{TUH EEG Corpus (TUEG) \cite{obeid2016temple}}
The Temple University Hospital EEG Corpus is a substantial collection of clinical EEG recordings, consisting of 26,846 recordings collected from 2002 to 2017, 
taking up 1.6 TB of storage. This dataset is the primary data source for developing and testing our self-supervised learning pipeline for EEG data analysis as it is used for pretraining. It also serves as the primary motivation as it is by far the largest public EEG dataset. Still, there is a vast difference in the recordings in terms of equipment used, task performed, length of recording, and quality of data.

\subsection{Motor Movement/Imagery Dataset (MMIDB) \cite{mmidb_dataset}}
The MMIDB dataset includes 64-channel EEG recordings from 109 subjects who performed motor and imagery tasks related to hand movements. We are excluding participants S088, S090, S092, and S100 due to missing data, resulting in 105 participants. We use sessions 3, 7, and 11 where a target appears on either the left or right side of the screen, and the subject opens and closes the corresponding fist.
This dataset is used for
downstream benchmarking.

\subsection{BrainCapture Bhutan Dataset v. 4.1 (BC Bhutan)}
The BC Bhutan dataset includes 27-channel EEG recordings from 133 subjects across Bhutan. The data was collected in collaboration with BrainCapture and is used for fine-tuning models. As part of routine EEG recordings, the data contain the 5-second annotated exercise windows \textit{Eye blinking}, \textit{Eye movement}, \textit{Eyes closed}, \textit{Eyes opened}, and \textit{Jaw clenching}. This dataset is used for
downstream benchmarking.

\subsection{BCI Challenge @ NER 2015 (BCI@NER) \cite{ner-bci-challenge}}
The dataset from the BCI Challenge at NER 2015 includes 56-channel EEG recordings from 16 subjects who participated in 5 sessions each, totalling 80 sessions. This dataset includes two classes for positive and negative feedback for P300 waves and is used for downstream benchmarking.

\section{Methods} \label{sec:theory}

The pipeline is built upon already existing preprocessing components that are implemented and available in various packages in Python. This section briefly introduces the methods we use in consecutive order as shown in Figure \ref{fig:pipeline}. The objective of SPEED is to preprocess data into a uniform format suitable for machine learning applications, ensuring high-quality data with consistency in channels and sample counts. We describe the general SPEED pipeline for self-supervised pretraining and address any changes for downstream datasets in a subsection afterwards.
\subsection{Initial Setup of preprocessing} %
The initial steps are simple yet essential. They consist of standardizing channel names, detecting channel types, dropping non-EEG channels, setting montage, and then dropping channels not in a known montage. We then segment the data into 60-second windows and omit the final window if it is less than 60 seconds in length. All subsequent procedures are carried out within these windows.
\subsection{Quality assessment}
The process of quality assessment involves quickly determining whether a window should be kept or discarded. This is not to waste unnecessary time on too poor quality windows. We chose to drop windows if there is an insufficient number of channels to interpolate the desired montage, either due to a lack of channels or because they are removed for being defective. Here we require at least half of the final number of channels to be present. Another reason is the excessive dropping of ICA components, leaving little of the original signal intact. Lastly, the data may be of inherently poor quality, either before or after processing.
Quality assessment begins with a straightforward bad channel detection using the PREP pipeline \cite{PREP} methods (excluding RANSAC), followed by the application of a notch filter to eliminate line noise and a combination of a 1 Hz high-pass and a 100 Hz low-pass filter. Subsequently, various quality measures are calculated, largely adopting the criteria recommended in the AutoMagic pipeline \cite{automagic}. These measures include the Ratio of Bad Channels (RBC), the Ratio of Data with Overall High Amplitude (OHA), the Ratio of Timepoints with High Variance (THV), and the Ratio of Channels with High Variance (CHV). We exclude any window where $OHA < 0.8$, $THV < 0.5$, $CHV < 0.5$, and $BCR < 0.8$. These thresholds were determined through manual reviews of over 100 EEG recordings.

\subsection{Iterative Zapline}
Next, we remove line noise using the Iterative Zapline Algorithm \cite{ZAPLINE2020}. The Zapline algorithm is designed to remove power line artifacts from multichannel electrophysiological data, such as EEG and MEG. It combines spectral and spatial filtering techniques to achieve effective noise removal while preserving the underlying signals \cite{ZAPLINE2020}. The line noise frequency, usually 50Hz in Afro-Eurasia or 60Hz in America, is manually set. The Python implementation MeegKit is applied \cite{ZAPLINE2020}.

\subsection{Bad Channel Detection}\label{sec:bad_channel_detection}
Bad channel detection is performed on a copy of the dataset, and the identified bad channels are afterwards removed from the original dataset. The objective of this process is to identify channels with low signal-to-noise ratios (SNR). Dropping these channels is essential as their presence negatively impacts interpolation, average referencing, and independent component analysis (ICA).

\quad The basis for bad channel detection lies in the Python implementation of the PREP pipeline. 
It initiates the detection by removing flat/NAN channels and low-frequency trends. It then identifies channels with extreme amplitudes (deviation criterion), abnormal high-frequency noise (noisiness criterion), poor correlation with other channels (correlation criterion), and weak predictability by other channels (predictability criterion). The latter involves assessing the correlation of each channel within a 5s EEG data window with its Random Sample Consensus (RANSAC) \cite{ransac} reconstruction (using 50 samples from a subset of 25\% of all channels). A channel is considered bad if its correlation is below 0.85 in more than 40\% of instances. Due to the RANSAC algorithm requiring significant resources, its inclusion is optional. The identified bad channels are then excluded from the dataset. Originally, the PREP pipeline included bad channel detection in its robust average referencing process, which iteratively detects and interpolates bad channels to ensure a reliable average reference. However, as we do not wish to interpolate every bad channel, particularly before ICA, and the high computational demands, this method is not used.

\subsection{Filtering \& Average Referencing}
We apply robust detrending, a 0.5 Hz high-pass filter, and a 100 Hz low-pass filter. We use robust detrending because we aim to eliminate linear trends, even though the specific frequency of the high-pass filter may vary as needed. Lastly, we apply average referencing to the remaining high-quality channels \cite{average_ref}, \cite{filters}.

\subsection{Independent Component Analysis}
We proceed to the Extended Infomax Independent Component Analysis (ICA) \cite{ExtendedInfomax} transformation, setting the number of components equal to the number of remaining channels. We classify the components using ICLabel \cite{PIONTONACHINI2019181} methods and exclude every independent component not categorized as \textit{brain} or \textit{other} and with certainty above 80\%. The signal is reconstructed from the remaining independent components. A further \textit{Bad Channel Detection} (see Section \ref{sec:bad_channel_detection}) is conducted on a duplicate of the dataset.

\quad Independent Component Analysis (ICA) stands out among various methods for isolating and eliminating artifacts and noise from data. It represents the most computationally intensive step in the process, yet could offer significant benefits, particularly through automation with ICLabel. Therefore we add an option to skip the ICA in the pipeline.

\vspace{0.5em}
\textbf{ICLabel:} The ICLabel project \cite{PIONTONACHINI2019181} is designed to automate the classification of electroencephalographic (EEG) independent components (ICs), aiming to facilitate the analysis of EEG studies with many subjects and enable the use of ICA decomposition automatically. This effort has produced the automated ICLabel classifier, trained on over 6,000 labelled ICs, and is currently leading in accuracy. It assigns ICs to seven categories: \textit{brain}, \textit{muscle artifact}, \textit{eye blink}, \textit{heart}, \textit{line noise}, \textit{channel noise}, and \textit{other}. The latter category serves as a general catch-all for ICs with indeterminate noise or those containing signals that ICA decomposition could not distinctly separate.

\subsection{Interpolation \& Resample}
Missing channel interpolation is crucial because it enables us to standardize EEG data from various sources and leverage information from additional channels instead of discarding them. We interpolate the channels that are missing from the final desired set, regardless of whether they were deemed bad or were initially missing. The interpolation is done with Legendre polynomial expansion via the minimum-norm method, as implemented in MNE Python. More information on these methods can be found in \cite{interpolation}. Afterwards, we remove any extra channels, reorder the final channels, and resample them to 256Hz.

\subsection{Downstream Data Preprocessing}
The downstream data preprocessing is done similarly to the pretraining data. The changes are highlighted in Figure \ref{fig:pipeline} in dotted lines. Those parts are omitted from the downstream processing.
After the initial setup, the downstream data is not split into 60-second windows. However, all data before the first and after the last event marker is removed to avoid unclean sections in the signal. All subsequent procedures are carried out on the cropped recording.
The quality assessment is skipped on downstream data because, in datasets intended for downstream analysis, the decision to drop a window is challenging, as it could increase the risk of Type I errors \cite{FalsePositive}. For example, in attempting to detect seizures, which are characterized by high-frequency noise, it is undesirable to discard windows with high variance. While this issue remains relevant in a self-supervised learning context with large-scale data, it becomes less prominent since the specifics of downstream analysis are not predetermined.
Other parts in the pipeline that require dropping bad windows are skipped because the downstream data is processed as one recording and not split into windows.

%% file: Sections/3_Results.tex
\section{Experiments}
\subsection{Preprocessing of the TUH EEG Corpus}
A significant contribution of this paper is the preprocessing of the Temple University Hospital EEG Corpus. SPEED is designed for multiprocessing and is executed in parallel across 30 cores on 2 x EPYC 7302 and 30 cores on 2 x EPYC 9124, each equipped with 128GB of RAM, completing the task in under a week. The preprocessing comprehensively logs each step, allowing for detailed inspection of each data window. This includes information on which windows are dropped, identification of bad channels, channels removed for not matching the montage setup, the labels and probability scores of independent components from ICLabel, components that are excluded, and extra channels that are dropped.

\subsection{Pretraining}
We use the BENDR model from Kostas et al. \cite{kostas2021bendr} for our experiments. This model is pretrained over 5 epochs using the preprocessed TUH EEG Corpus with (1) SPEED, as depicted in Figure \ref{fig:pipeline}, (2) SPEED w/ ICA, also shown in Figure \ref{fig:pipeline}, and (3) Baseline, as employed in Kostas et al. \cite{kostas2021bendr}, which involves only resampling and zeroing out missing channels. For validation, we similarly preprocess the three smaller downstream datasets (MMIDB, BC Bhutan, and BCI@NER) in 60-second windows. The model with the lowest validation loss from this phase is selected for subsequent probing.
\subsection{Probing}
We initiate probing by incorporating the pre-trained encoder layer from the model, adding linear layers, and locking the pre-trained encoder layer. This procedure aims to assess whether the pretrained latent space provides an improved representation of EEG data when processed using different preprocessing pipelines. The models are probed using the labelled downstream datasets MMIDB, BC Bhutan, and BCI@NER, which are preprocessed similarly and segmented into 5-second windows.

\quad We implement a 10-fold cross-validation with subject-based splits to ensure that data from the same individual does not overlap across the training, validation, and test sets. In each cross-validation cycle, the designated test fold is split further into two distinct datasets: one for validation and one for testing. The test dataset is used for ongoing evaluation and model selection throughout the training phase. The validation dataset, on the other hand, is used exclusively for the final evaluation of the model’s performance post-training. This method guarantees that our model evaluation is thorough and accurately represents the model's effectiveness with new, unseen data. This configuration is repeated five times resulting in 50 performance estimates per experimental setup for each downstream dataset.

\section{Results} \label{sec:results}
\subsection{Preprocessing}
\begin{figure}[H]
\centering
\includegraphics[width=1.0\linewidth]{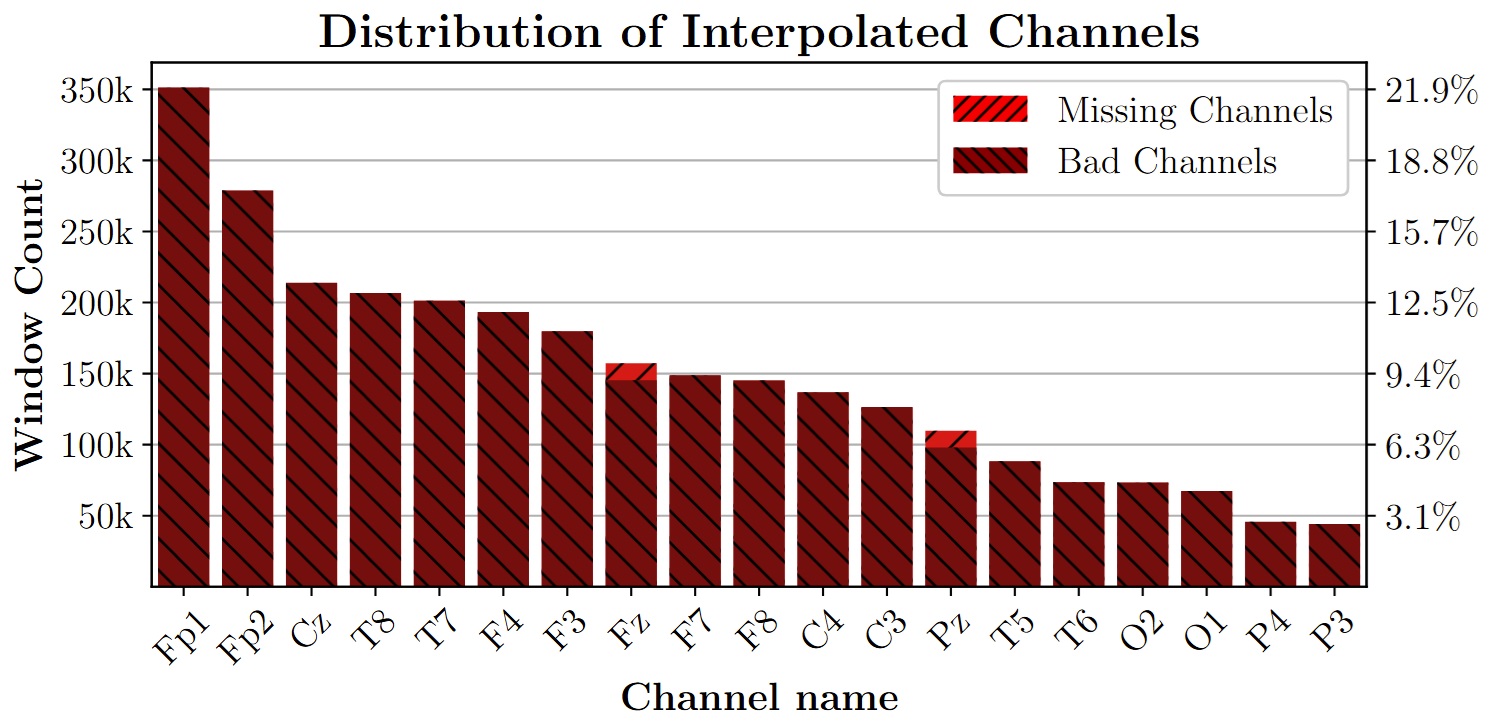}
\vspace{-18pt}
\caption{Distribution of interpolated channels during preprocessing with SPEED and SPEED w/ ICA. Most of the channels are interpolated for being detected as \textit{bad} while a few instances for \textit{Fz} and \textit{Pz} were originally missing.}
\label{preprocess:fig:interpolated}
\end{figure}

\begin{figure}[H]
\centering
\includegraphics[width=1.0\linewidth]{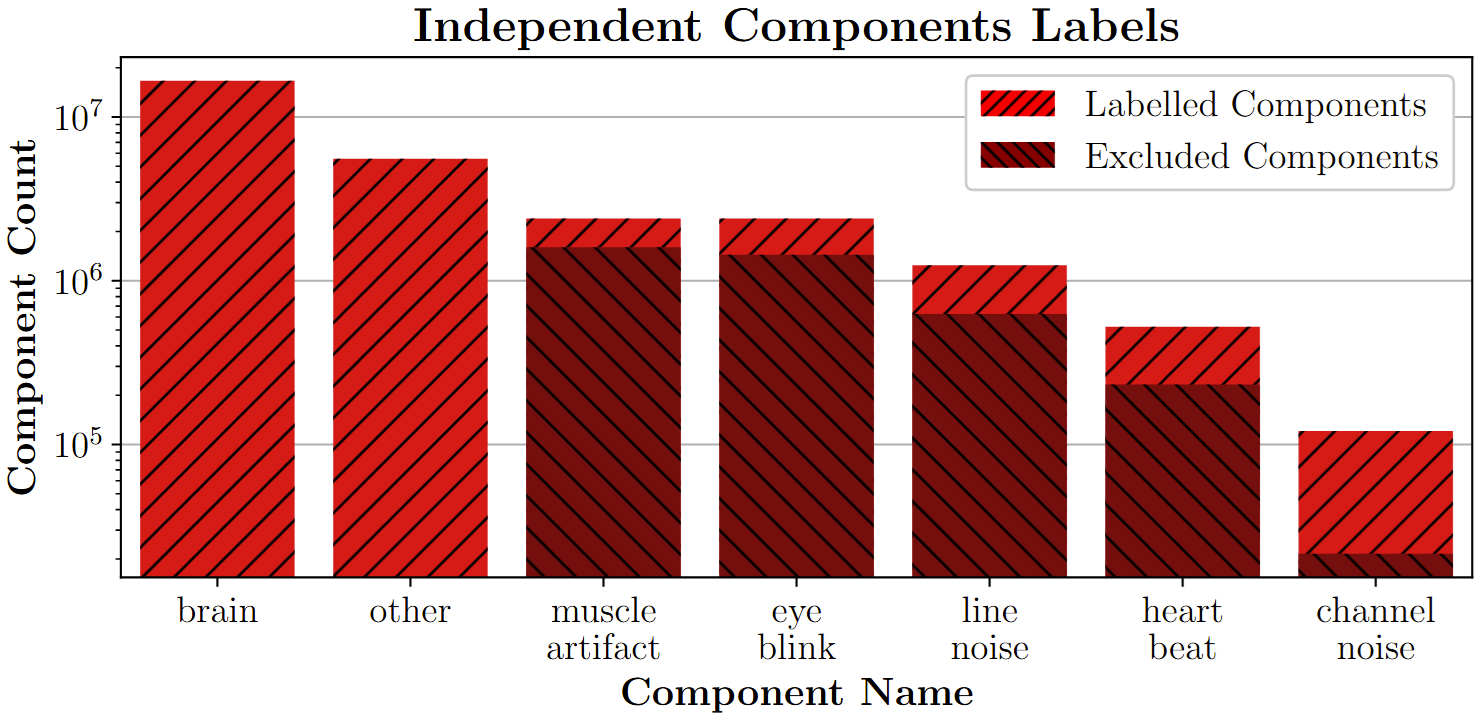}
\vspace{-18pt}
\caption{Distribution of the independent component (ICs) classification by ICLabel during preprocessing with the SPEED pipeline on The TUH EEG Corpus. Most of the ICs are classified as \textit{brain} and \textit{other} as expected.}
\label{preprocess:fig_independent}
\end{figure}

In Figure \ref{preprocess:fig:interpolated}, we identify which channels are interpolated at the final step due to being initially missing or removed for poor quality. 
%
%
\quad The independent component analysis, crucial for reducing data loss, depends on ICLabel's accuracy. As illustrated in Figure \ref{preprocess:fig_independent}, the majority of components are categorized as \textit{brain} and \textit{others}. Components identified as \textit{muscle artifact}, \textit{eye blink}, \textit{line noise}, and \textit{heart beat} typically demonstrate a certainty above 80\% and are excluded. The component \textit{channel noise} is frequently detected, however, due to generally low classification certainty, few are removed.

\subsection{UMAP Embedding Analysis}

\begin{figure}
\centering
\begin{subfigure}[b]{0.5\textwidth}
   \includegraphics[width=1\linewidth]{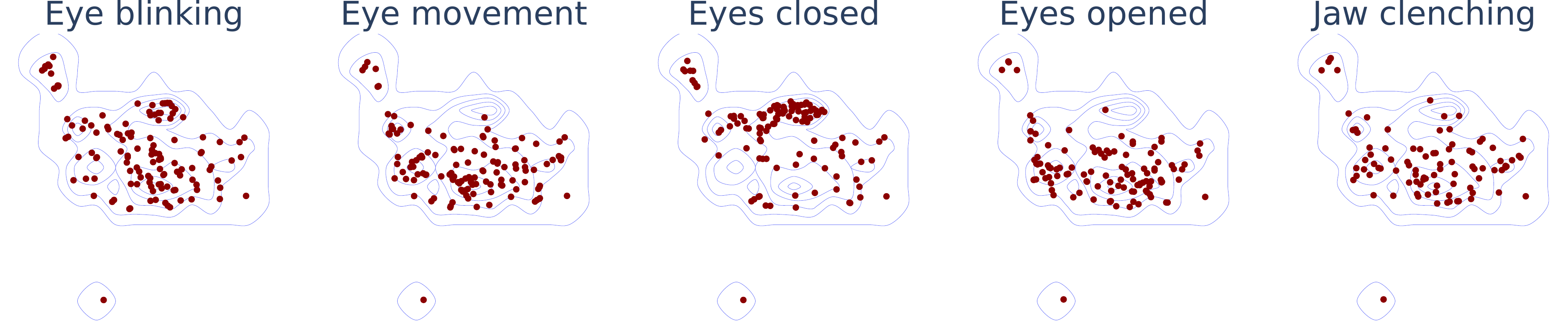}
   \caption{Preprocessed with Baseline Pipeline}
   \label{fig:umap1} 
\end{subfigure}

\begin{subfigure}[b]{0.5\textwidth}
    \vspace{1em}
   \includegraphics[width=1\linewidth]{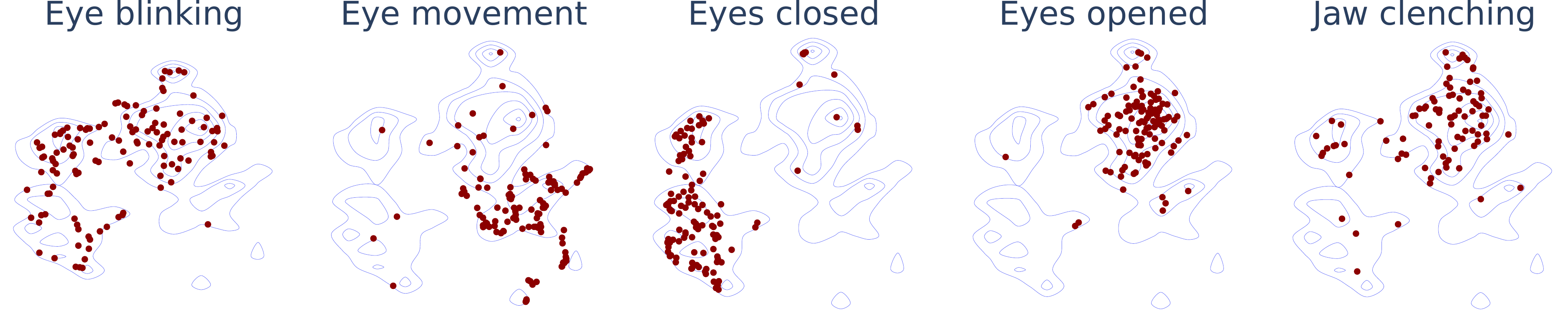}
   \caption{Preprocessed with SPEED Pipeline}
   \label{fig:umap2}
\end{subfigure}
\caption[]{UMAP embeddings of the Bhutan dataset. The subplots show the distribution of the 5 artifact classes as scatterplots. The density of the whole dataset is represented by the contour plot. The representations based on SPEED show better alignment with the ground truth labels.
}
\label{fig:umap}
\vspace{-2em}
\end{figure}

Figure \ref{fig:umap} motivates the application of preprocessing. We use UMAP to visualize the embedding of the BrainCapture Bhutan dataset, obtained without reference to the provided labels. We compare data preprocessed with (\ref{fig:umap1}) the Baseline and (\ref{fig:umap2}) the SPEED pipeline. Significant improvements in class separation are noticeable when using the preprocessing approach. In addition to visually inspecting the label clusters, we use Fisher's Linear Discriminant (FLD) score \cite{li2014fisher} to quantify the ratio of between-class variance to within-class variance. With an FLD-score of 0.0403, we find a much stronger separation between (\ref{fig:umap1}) data preprocessed with SPEED, compared to (\ref{fig:umap2}) data preprocessed with the baseline pipeline, which has an FLD-score of 0.0009 (indicating little to no association).

\subsection{Pretraining}
\begin{figure}[h]
\centering
\includegraphics[width=1.0\linewidth]{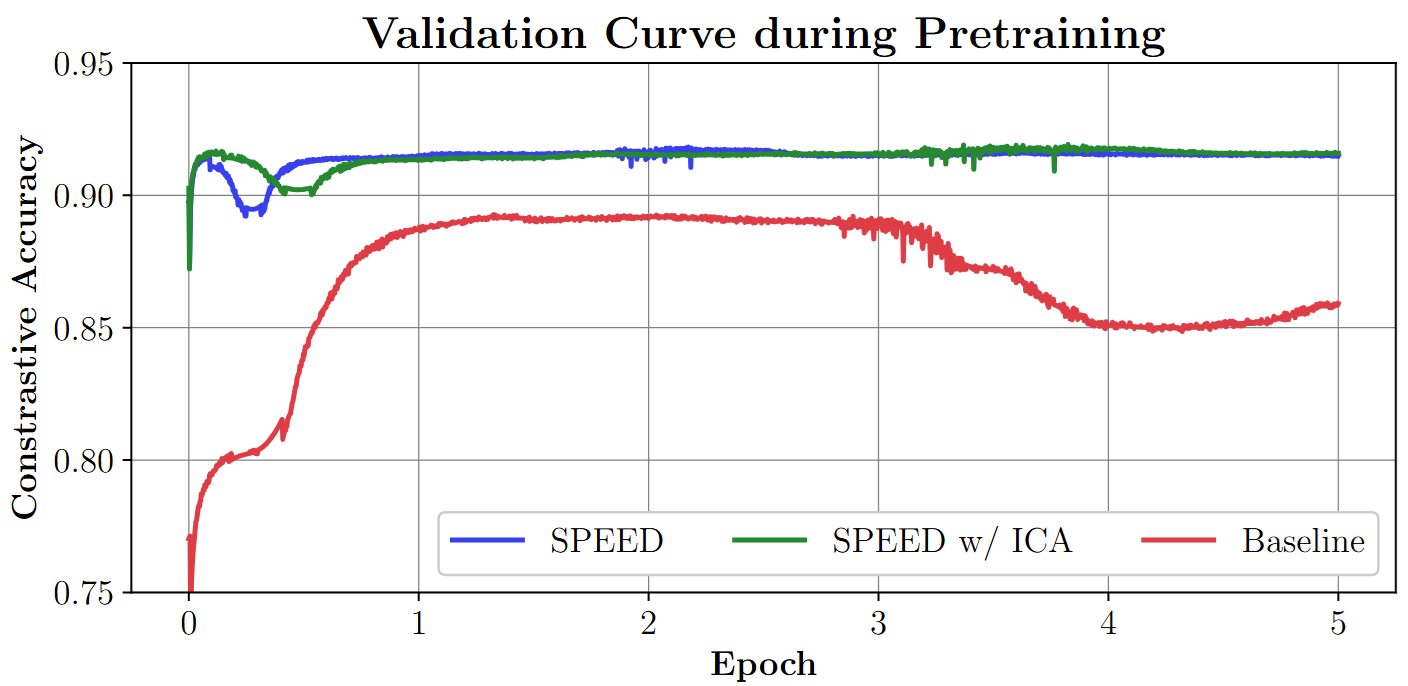}
\vspace{-18pt}
\caption{Validation contrastive accuracy during pretraining for three different versions of preprocessed datasets; SPEED, SPEED w/ ICA, and Baseline. The models with SPEED and SPEED w/ ICA offer more stable training and achieve higher scores.}
\label{fig:pretrain}
\end{figure}

In Figure \ref{fig:pretrain}, the validation contrastive accuracy curves are displayed for the pretraining using three different preprocessed versions of The TUH EEG Corpus. Each model increases rapidly during the initial iterations but experiences a drop in accuracy halfway through the first epoch. However, the model using the Baseline data takes several epochs to reach its peak, while the other models only require about one epoch. Additionally, the Baseline model achieves a maximum accuracy of 89\%, whereas the other models reach 92\% accuracy. Lastly, the Baseline model begins to overfit after approximately four epochs, while the other models maintain a consistent, slight increase in validation accuracy. The use of SPEED and SPEED w/ ICA offers a more stable pretraining and a final higher contrastive accuracy on the validation datasets.
\subsection{Probing of the model}
Table \ref{results:probing} presents the results of probing the model. It clearly shows that both SPEED and SPEED w/ ICA during pretraining enhance the model's downstream accuracy compared to the baseline. Specifically, when compared to Baseline, using SPEED for pretraining, results in an accuracy increase of 9\% for MMIDB, 7\% for BrainCapture Bhutan, and 6\% for BCI@NER.
\begin{table}[h]
\centering
\resizebox{\columnwidth}{!}{
\begin{tabular}{|c|c|c|c|}
\hline
\begin{tabular}{@{}c@{}}Pretrain $\rightarrow$  \\ $\downarrow$ Downstream\end{tabular} &  \textbf{SPEED} &  \textbf{SPEED w/ ICA} & \textbf{Baseline}  \\ \hline
\textbf{MMIDB} & \multicolumn{3}{c|}{}\\ \hline
SPEED & $\mathbf{0.73 \pm 0.01}$ & $0.71 \pm 0.01$ & $0.64 \pm 0.01$             \\ \hline
SPEED w/ ICA & $\mathbf{0.66 \pm 0.01}$ & $0.65 \pm 0.02$ & $0.59 \pm 0.02$            \\ \hline
Baseline & $\mathbf{0.76 \pm 0.01}$ & $0.74 \pm 0.01$ & $0.70 \pm 0.01$             \\ \hline
\textbf{BC Bhutan} &\multicolumn{3}{c|}{} \\ \hline
SPEED & $\mathbf{0.48 \pm 0.02}$ & $0.43 \pm 0.02$ & $0.41 \pm 0.02$             \\ \hline
SPEED w/ ICA & $0.31 \pm 0.03$ & $\mathbf{0.34 \pm 0.03}$ & $0.32 \pm 0.03$              \\ \hline
Baseline & $\mathbf{0.36 \pm 0.03}$ & $0.34 \pm 0.03$ & $0.33 \pm 0.02$             \\ \hline
\textbf{BCI@NER} & \multicolumn{3}{c|}{}\\ \hline
SPEED & $\mathbf{0.68 \pm 0.03}$ & $0.67 \pm 0.03$ & $0.62 \pm 0.03$             \\ \hline
SPEED w/ ICA & $\mathbf{0.67 \pm 0.02}$ & $\mathbf{0.67 \pm 0.02}$ & $0.64 \pm 0.02$              \\ \hline
Baseline & $\mathbf{0.70 \pm 0.03}$ & $\mathbf{0.70 \pm 0.03}$ & $0.65 \pm 0.03$             \\ \hline
\end{tabular}
}
\caption{Accuracy with std. dev. from the probing results, obtained through five repetitions of 10-fold cross-validation. The columns represent the three different preprocessing versions of the TUH EEG Corpus: SPEED, SPEED w/ ICA, and Baseline, on which the model is pre-trained. The rows show the preprocessing applied to the downstream dataset. This is assessed across three different downstream datasets: MMIDB, BC Bhutan, and BCI@NER. An improvement is observed with the use of SPEED preprocessing. The best performance per row is boldfaced.}
\label{results:probing}
\end{table}
However, it is also clear that while SPEED w/ ICA during pretraining is advantageous compared to the Baseline, it does not outperform SPEED. On the MMIDB and BC Bhutan datasets SPEED w/ ICA achieves a significantly lower accuracy. A plausible explanation for this could be removing muscle and eye artifacts during the preprocessing phase. For the BrainCapture Bhutan dataset, these artifacts are directly related to what we are attempting to classify, and for MMIDB, they could potentially be used to enhance classification indirectly.
%
%

%% file: Sections/4_Conclusion.tex
 \section{Discussion}
With our pipeline, we address a major challenge of large-scale EEG data preprocessing pipelines. However, a major consideration is whether to incorporate ICA with ICLabel. This technique is well-established in EEG preprocessing and analysis and is part of the AutoMagic \cite{automagic} pipeline. Our analysis indicates that including ICA does not significantly improve the pretraining performance. Rather, examining the probing results slightly reduces the downstream accuracy when used in pretraining and significantly when applied to downstream datasets. This suggests that the model might either miss artifacts in SPEED, perhaps use them in classification or that ICLabel incorrectly classifies independent components, leading to the removal of significant data.

\quad For downstream datasets, more careful preprocessing is often required, which is not the focus of this paper. These datasets are typically smaller and of higher quality, making them more suitable for detailed analysis. Given that both the MMIDB and BrainCapture Bhutan classification tasks involve muscle movement, it is unsurprising that removing the \textit{muscle artifact} and \textit{eye blink} ICs diminishes accuracy.

\quad Another point of interest is that we are fine-tuning the model using the same datasets designated for validation during pretraining, which could introduce bias. We adopt this approach as it aligns with the methodology used in the paper by Kostas et al. \cite{kostas2021bendr}. During pretraining, we use 60-second windows and all available data, both annotated and unannotated, with the latter making up the majority of the dataset. Therefore, this potential bias is not a major concern.

\section{Conclusion} \label{sec:conclusion}
The 'industry standard' in machine learning is to let complex deep learning models handle massive datasets with limited preprocessing, especially with self-supervised learning. For electroencephalogram (EEG) data, Kostas et al.\ pretrained a complex transformer model on the massive Temple University Hospital (TUH) EEG Corpus with rather rudimentary preprocessing. However, the TUH EEG Corpus presents significant challenges due to its variability in signal-to-noise, equipment used, length of recording, and more. We introduce an efficient preprocessing pipeline designed to handle variability within one or more EEG datasets and combine them into a single preprocessed dataset useful for self-supervised learning applications. The Python-based pipeline improves stability, convergence, and contrastive accuracy during pretraining and produces a more suitable latent space for downstream classification tasks. More specifically, our probing results for downstream accuracy show a significant improvement for several downstream classification tasks when pretraining using our preprocessing pipeline compared to a simple baseline preprocessing pipeline. Besides the plug-and-play preprocessing pipeline, we also present tools for reproducible preprocessing of the TUH EEG Corpus ready for future development of pretrained EEG foundation models. In conclusion, our results form evidence that physically motivated preprocessing is useful for self-supervised learning of EEG representations.
